\documentclass[letterpaper,10pt]{article} 

\usepackage{opticameet3} %

\newcommand\authormark[1]{\textsuperscript{#1}}

\usepackage{amsmath,amssymb}
\usepackage[colorlinks=true,bookmarks=false,citecolor=blue,urlcolor=blue]{hyperref} 
\usepackage[font={footnotesize}]{caption}

\usepackage[nohyperlinks,nolist]{acronym}
\usepackage{booktabs}
\usepackage{multirow}
\usepackage{pgfplots}
\usepgfplotslibrary{groupplots}  %
\pgfplotsset{compat=newest}
\usepackage{wrapfig}
\usepackage{caption}
\usepackage{bm}
\usetikzlibrary{positioning, arrows.meta, shapes.geometric}

\newcommand{\vect}[1]{\bm{#1}}
\newcommand{\bs}[1]{\boldsymbol{#1}}

\newcommand{\cout}{{C_\text{out}}}

\newlength{\ww}
\newlength{\pw}%
\newlength{\ph}%
\newlength{\hs}%
\newlength{\vs}%
\definecolor{KITgreen}{rgb}{0,.59,.51}
\colorlet{kit-green100}{KITgreen!100}
\colorlet{kit-green70}{KITgreen!70}
\colorlet{kit-green50}{KITgreen!50}
\colorlet{kit-green30}{KITgreen!30}
\colorlet{kit-green15}{KITgreen!15}

\definecolor{KITpalegreen}{RGB}{130,190,60}
\colorlet{kit-palegreen100}{KITpalegreen}
\colorlet{kit-palegreen70}{KITpalegreen!70}
\colorlet{kit-palegreen50}{KITpalegreen!50}
\colorlet{kit-palegreen30}{KITpalegreen!30}
\colorlet{kit-palegreen15}{KITpalegreen!15}

\definecolor{KITblue}{rgb}{.27, .39, .66}
\colorlet{kit-blue100}{KITblue!100}
\colorlet{kit-blue70}{KITblue!70}
\colorlet{kit-blue50}{KITblue!50}
\colorlet{kit-blue30}{KITblue!30}
\colorlet{kit-blue15}{KITblue!15}
 
\definecolor{KITyellow}{rgb}{.98, .89, 0}
\colorlet{kit-yellow100}{KITyellow!100}
\colorlet{kit-yellow70}{KITyellow!70}
\colorlet{kit-yellow50}{KITyellow!50}
\colorlet{kit-yellow30}{KITyellow!30}
\colorlet{kit-yellow15}{KITyellow!15}
 
\definecolor{KITorange}{rgb}{.87, .60, .10}
\colorlet{kit-orange100}{KITorange!100}
\colorlet{kit-orange70}{KITorange!70}
\colorlet{kit-orange50}{KITorange!50}
\colorlet{kit-orange30}{KITorange!30}
\colorlet{kit-orange15}{KITorange!15}
 
\definecolor{KITred}{rgb}{.63, .13, .13}
\colorlet{kit-red100}{KITred!100}
\colorlet{kit-red70}{KITred!70}
\colorlet{kit-red50}{KITred!50}
\colorlet{kit-red30}{KITred!30}
\colorlet{kit-red15}{KITred!15}
 
\definecolor{KITpurple}{RGB}{160, 0, 120}
\colorlet{kit-purple100}{KITpurple}
\colorlet{kit-purple70}{KITpurple!70}
\colorlet{kit-purple50}{KITpurple!50}
\colorlet{kit-purple30}{KITpurple!30}
\colorlet{kit-purple15}{KITpurple!15}
 
\definecolor{KITcyanblue}{RGB}{80, 170, 230}
\colorlet{kit-cyanblue100}{KITcyanblue}
\colorlet{kit-cyanblue70}{KITcyanblue!70}
\colorlet{kit-cyanblue50}{KITcyanblue!50}
\colorlet{kit-cyanblue30}{KITcyanblue!30}
\colorlet{kit-cyanblue15}{KITcyanblue!15}

\acrodef{MLP}[MLP]{multi-layer perceptron}
\acrodef{KAN}[KAN]{Kolmogorov-Arnold network}
\acrodef{CNN}[CNN]{convolutional neural network}
\acrodef{ROP}[ROP]{received optical power}
\acrodef{sps}[sps]{samples per symbol}
\acrodef{MSE}[MSE]{mean squared error}
\acrodef{LR}[LR]{learning rate}
\acrodef{rvms}[rvms]{real-valued multiplications per symbol}
\acrodef{BER}[BER]{bit error rate}
\acrodef{IM/DD}[IM/DD]{intensity-modulation and direct-detection}
\acrodef{GRU}[GRU]{gated recurrent unit}
\acrodef{ANN}[ANN]{artificial neural network}
\acrodef{FIR}[FIR]{finite impulse response}
\acrodef{LUT}[LUT]{lookup table}
\acrodef{MSB}[MSB]{most significant bit}

\acrodef{CD}[CD]{chromatic dispersion}

\acrodef{DFB}[DFB]{distributed-feedback laser}
\acrodef{DSP}[DSP]{digital signal processing}

\acrodef{EAM}[EAM]{electro-absorption modulator}

\acrodef{FPGA}[FPGA]{field-programmable gate array}

\acrodef{SSFM}[SSFM]{standard single-mode fiber}
\acrodef{SOA}[SOA]{semiconductor optical amplifier}
\acrodef{SNR}[SNR]{signal-to-noise ratio}

\acrodef{NRZ}[NRZ]{non-return-to-zero}

\acrodef{ONU}[ONU]{optical network unit}
\acrodef{OLT}[OLT]{optical line terminal}
\acrodef{OOK}[OOK]{on-off-keying}

\acrodef{PAM4}[PAM4]{4-ary pulse amplitude modulation}
\acrodef{PON}[PON]{passive optical network}

\acrodef{VOA}[VOA]{variable optical attenuator}

\begin{document}

\title{Non-linear Equalization in 112 Gb/s PONs Using Kolmogorov-Arnold Networks}

\author{Rodrigo Fischer,\authormark{1,*} Patrick Matalla,\authormark{2} Sebastian Randel,\authormark{2} and Laurent Schmalen\authormark{1}}

\address{\authormark{1}Communications Engineering Lab (CEL), Karlsruhe Institute of Technology, 76131 Karlsruhe, Germany\\ 
\authormark{2}Institute of Photonics and Quantum Electronics (IPQ), Karlsruhe Institute of Technology, 76131 Karlsruhe, Germany}

\email{\authormark{*}\texttt{rodrigo.fischer@kit.edu}} %

\begin{abstract}
We investigate \acp{KAN} for non-linear equalization of 112~Gb/s PAM4  \acp{PON}. Using pruning and extensive hyperparameter search, we outperform linear equalizers and convolutional neural networks at low computational complexity.
\end{abstract}
\vspace{0pt}
\section{Introduction}

\acresetall

\Acp{PON} are optical access networks whose branching points consist entirely of passive elements, enabling a cost-effective implementation. They currently serve the majority of fiber broadband subscribers worldwide and an ongoing demand for bandwidth has led to recent standardization efforts that enabled 50~Gb/s line rate transmission \cite{50GPON}, while the research community is investigating the technologies that will enable \acp{PON} beyond 100~Gb/s \cite{bonk24}. One possibility for achieving 100~Gb/s is the use of higher-order modulation formats in \ac{IM/DD} links. However, this comes at the cost of an increased \ac{SNR} requirement and lower tolerance to non-linearities in the channel. 
In a \ac{PON}, the \acp{SOA} used to improve the receiver sensitivity suffer from non-linear gain saturation and the \ac{EAM} responsible for modulating the intensity of the optical signal has a non-linear transfer function. Additionally, when using \ac{IM/DD}, the \ac{CD} corresponds to a non-linear channel effect. This motivates the use of non-linear equalizers at the receiver, specially with low computational complexity as costs in access networks are crucial. Previous works investigated \ac{GRU}-based \cite{murphy2023} equalizers as well as \acp{CNN}-based equalizers \cite{lau24}, \cite{ecocCNN}.

In this paper, we investigate \acp{KAN}, a novel type of neural network architecture able to learn non-linear activation functions, as equalizers in \acp{PON}. They were first used in applications where the analysis of the learned activation functions was used to infer information about the training data, helping scientists discover symbolic formulas or identify important features in mathematical or physical problems \cite{liu2024kan}. \Acp{KAN} exhibit two main features that motivate their use as non-linear equalizers.
First, smaller \acp{KAN} can reportedly achieve comparable performance as traditional \ac{MLP} networks \cite{liu2024kan}. With this, we expect that the \ac{KAN} non-linear activation functions can learn how to invert the non-linear channel more easily than \ac{MLP} networks with ReLU activation functions. A second aspect is that there are a few simple strategies that we can use to obtain a multiplier-free implementation of the \ac{KAN}.

In this work, we propose to use \acp{KAN} as non-linear equalizers for \acp{PON}. We compare them with \acp{CNN} and equalizers based on \ac{FIR} linear filters with the same computational complexity. We also highlight further characteristics of \acp{KAN} that make them good candidates for non-linear low-complexity \mbox{equalizers}.

\section{Kolmogorov-Arnold Networks (KANs)}

\begin{wrapfigure}[12]{r}{0.5\textwidth}
    \centering
    \input{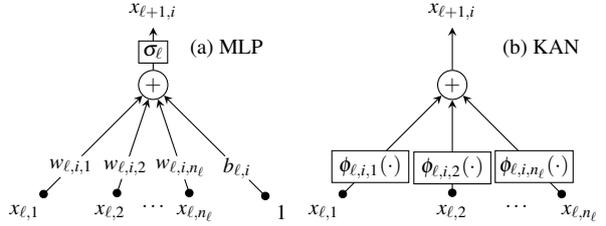}
    \captionsetup{width=\linewidth}
    \caption{Diagram of a single neuron in (a) a multi-layer perceptron (MLP) network and (b) a Kolmogorov-Arnold network (KAN) network, where boxes represent 1D functions.}
    \label{fig:diagram}
\end{wrapfigure}

In a conventional \ac{MLP} network, a layer consists of a linear transform followed by a non-linear activation function, whose operation can be written as $\vect{x}_{\ell+1}=\sigma_\ell(\vect{W}_\ell \vect{x}_\ell + \vect{b}_\ell)$, where $\ell \in \{1, ..., L\}$ is the layer index, $\vect{W}_\ell$ is the weight matrix, $\vect{b}_\ell$ is the the bias vector, $\vect{x}_\ell$ of size $n_\ell$ are the inputs and outputs of the $\ell$-th and $(\ell -1)$-th layers, and $\sigma_\ell$ is the activation function, which can be chosen as ReLU, ELU, sigmoid, etc. \cite{Goodfellow-et-al-2016}, as depicted in Fig.~\ref{fig:diagram}\,(a).

\acp{KAN} propose a different layer operation, where the multiplications of the weights by the inputs are replaced by trainable 1D functions $\phi_{\ell, i, j}(x_{\ell, j})$ \cite{liu2024kan}, as shown in Fig. \ref{fig:diagram}\,(b), where each entry $x_{\ell+1, i}$ of the output vector is obtained by computing \mbox{$x_{\ell+1, i} = \sum_{j\in\{1,...,n_\ell\}}\phi_{\ell, i, j}(x_{\ell, j})$}. This can be conveniently represented by a matrix-function notation $\vect{x}_{\ell+1} = \bs{\Phi}_\ell(\vect{x}_\ell)$, where $\bs{\Phi}_\ell = \{\phi_{\ell, i, j}\}$ contains the functions of layer $\ell$, which are applied to the entries of $\vect{x}_\ell$ and summed in a matrix-vector multiplication fashion. With multiple layers, we have \mbox{$\text{KAN}(\vect{x})=\bs{\Phi}_L(\bs{\Phi}_{L-1}(\cdots\bs{\Phi}_1(\vect{x})))$}, which enables the network to learn compositional relationships involving the data features.

A straightforward way to implement trainable 1D functions is to use B-splines \cite{liu2024kan}; we use linear B-splines centered at $G$ grid points located at $\{-4, \ldots, 4-t, 4\}$, with $t = 8 / (G - 1)$, which results in activation functions of the form \mbox{$\phi_{\ell, i, j}(x) = \sum_{k=0}^{G-1} a_{\ell, i, j, k} \text{tri} (x + 4 - kt)$}, where $a_{\ell, i, j, k} \in \mathbb{R}$ are the trainable parameters and $\text{tri}(x)$ is the triangular function with unit height spanning from $x=-t$ to $x=t$. This corresponds to a piece-wise linear function that interpolates the values $a_{\ell, i, j, k}$ for $k=0,\ldots,G-1$ between the grid points. This can be implemented in hardware by first determining where the input $x$ lies on the grid, followed by a single multiplication by a slope and a summation with an offset. The slopes and offsets can be stored in a \ac{LUT} with $G-1$ entries.

\section{Experimental Results and Discussion}

\begin{figure}[b]
    \centering
    \input{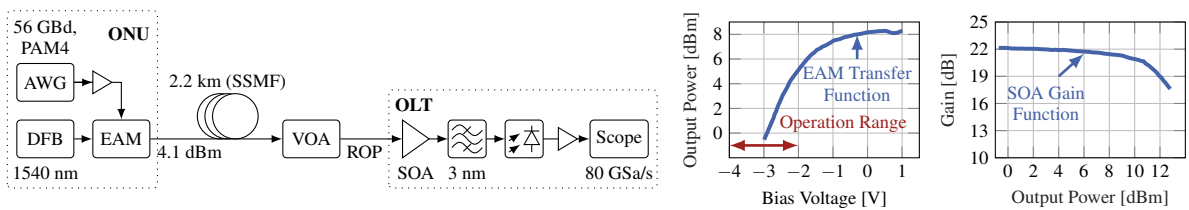}
    \caption{Experimental setup for a 112~Gb/s (56 GBd, \ac{PAM4}) PON upstream through a \ac{SSFM} in the C-band. In the right, the non-linear \ac{EAM} transfer function and \ac{SOA} gain function is shown.}
    \label{fig:experiment}
\end{figure}

We validate the novel \ac{KAN}-based equalizer in a 112~Gb/s \ac{PON} upstream transmission in the C-band, as depicted in Fig.~\ref{fig:experiment}. The setup is similar to the one reported in \cite{lau24} but with some improvements regarding device losses and \ac{EAM} operation point. At the \ac{ONU} side, we generate $2^{19}$ symbols in a \ac{FPGA} and use a Keysight USPA DAC3 signal converter to generate a 56~GBd \ac{NRZ} \ac{PAM4} signal with a peak-to-peak voltage of 2~V. The signal modulates the optical light emitted by a \ac{DFB} laser at 1540~nm in a low-cost \ac{EAM} modulator, which will distort the signal due to its non-linear transfer function (see Fig.~\ref{fig:experiment}, right). Afterwards, the signal is launched with an optical power of 4.1~dBm into a 2.2~km-long fiber, resulting in a total accumulated \ac{CD} of about $35.9\,\text{ps}\cdot\text{nm}^{-1}$ for a fiber-dependent \ac{CD} coefficient of $16.3\,\text{ps}\cdot\text{nm}^{-1}\cdot\text{km}^{-1}$. Using a \ac{VOA}, we can adjust the \ac{ROP}. At the \ac{OLT} receiver, an \ac{SOA} with 3~nm optical bandpass filter is used due to the increased \ac{SNR} requirements of \ac{PAM4} signals. Note that the \ac{SOA} will add non-linear distortions caused by gain saturation, as shown in Fig.~\ref{fig:experiment}.
The optical signal is detected using a 40~GHz-photodiode with conventional amplifier. Note, that further improvements in receiver sensitivity can be achieved using an avalanche photodiodes with transimpedance amplifier. The electrical signal is sampled using a 33~GHz-real-time oscilloscope at 80~GSa/s. In offline \ac{DSP}, the signal is resampled to twofold oversampling and synchronized using a blind feedforward clock recovery offering fast synchronization for \acp{PON} \cite{matalla24} before the \ac{KAN} equalizer is applied. 

For a \ac{ROP} ranging from $-30$\,dBm to 2\,dBm with a step size of 1\,dBm, we collected 30 sequences per \ac{ROP} with a length of $2.8\cdot 10^6$ PAM4 symbols at 2 \ac{sps}. One training iteration, where the network weights are updated, consisted of feeding 360 symbols in a convolution-manner through the equalizers for a total of 7600 iterations. Furthermore, we used the last 200 blocks of size 360 symbols of each sequence for testing, performed at the end of each training iteration. For all the models, we use the \ac{MSE} loss and a \ac{LR} scheduler that multiplies the \ac{LR} by a factor of $\alpha_{\text{LR}}=0.4$ when the training loss reaches a plateau. We also used $L_1$ regularization of the weights with a factor of $5\cdot 10^{-3}$, which enforces sparsity of the weights.

We considered 2-layer networks \mbox{KAN-2} and \mbox{CNN-2}, as well as 1-layer network KAN-1, and \ac{FIR} filters with different amount of taps. For the multi-layer networks, we perform layer-wise convolution of the inputs with $\cout$ kernels (filters) of size $K$ taps, moving the filter $S$ samples at a time (stride). Table~\ref{tab:search_space} shows an overview of the search space for the network architectures. Higher strides result in lower complexity per-symbol \cite{lau24}. We also fix $\cout$ in the last layer as $1/\text{sps}\cdot\prod_{\ell=1}^L S_\ell$, where $S_\ell$ is the stride of the $\ell$-th layer, which guarantees that symbols enter and leave the equalizer at equal rate. Furthermore, we searched the \ac{LR} over \mbox{$\{1, 1.77, 3.16, 5.62\}\cdot 10^{-3}$}. Pruning was performed by forcing the weights whose magnitude lies below a threshold w.r.t. the maximum weight magnitude for each layer to zero. For the \acp{KAN} we used $\sum_{k=0}^{G-1} |a_{\ell, i, j, k}|$ as a replacement for the magnitude of the weights. We used the thresholds $\{0, 1.25, 2.5, 5, 10, 15, \ldots, 95\}\%$ and we performed another round of training on the same sequences to adjust the remaining weights. In order to obtain the Pareto fronts, we used a subset of 8 sequences at $\text{ROP}=-2\,\text{dBm}$ and for each sequence we averaged the test \ac{BER} over the last 10 training iterations. To get a final metric, we took the maximum over the 8 sequences, which we denote by $\text{max}\left\{\overline{\text{BER}}\right\}$. By using the maximum metric, we hope to avoid choosing \ac{KAN} and \ac{CNN} architectures that may have a good median or mean performance, but that occasionally don't converge. We measure complexity in terms of the number of \ac{rvms}.

\begin{figure}[h]
    \centering
    \input{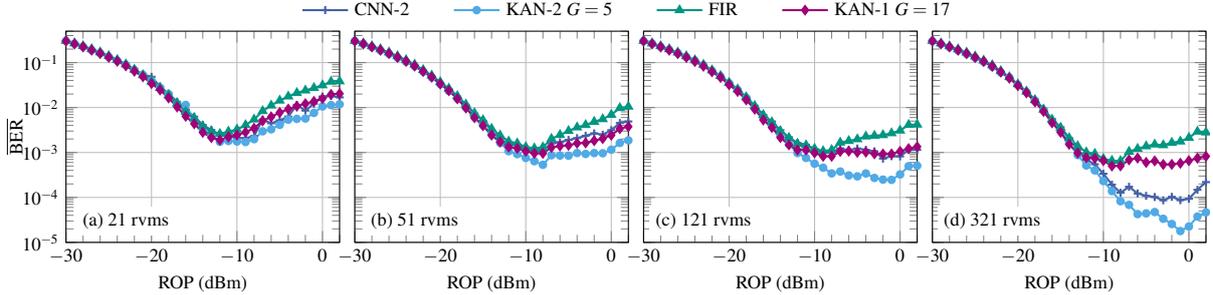}
    \captionsetup{width=\linewidth}
    \caption{Mean $\overline{\text{BER}}$ for a fiber length of 2.2 km (C-band) for models taken from the pruned Pareto curve with different complexities of 21 rvms (a), 51 rvms (b), 121 rvms (c) and 321 rvms (d).}
    \label{fig:deployment}
\end{figure}

The Pareto fronts can be seen in Fig.~\ref{fig:paretos} for the unpruned (a) and pruned (b) architectures. We observe a large gap between the \acp{CNN} and \acp{KAN} for the unpruned networks and a smaller gap for the pruned networks, which seem to indicate that the \acp{KAN} are more efficient in using the available weights than the \acp{CNN}, but once we remove the unused weights from the \ac{CNN}, we have a Pareto front closer to that of the \acp{KAN}. We show the performance of the models that lay on the pruned Pareto fronts at varying complexities on the entire measured \ac{ROP} range in Fig.~\ref{fig:deployment}. The KAN-2 outperforms all the other models, but notably, the KAN-1 is able to have similar performance to the CNN-2 up until 121 rvms. Also, the gap between the \ac{FIR} linear equalizers and the non-linear equalizers increase as the computational complexity increase.

\begin{figure}[t]
    \begin{minipage}[c]{0.36\textwidth}
        \setlength{\tabcolsep}{2pt}
        \scriptsize
        \centering
        \captionof{table}{Overview of the Search Space}
        \begin{tabular}{@{}clccccc@{}}
            \toprule
            \multicolumn{2}{c}{Net}                                                                 & Layer   & ${C_\text{out}}$    & $K$                                                                & $S$           & $G$          \\ \midrule
            \multicolumn{2}{c}{\multirow{2}{*}{\begin{tabular}[c]{@{}c@{}}KAN-2/\\ CNN-2\end{tabular}}} & Layer 1 & $\{2, 4, 8\}$       & $\{8, 32, 64\}$                                                    & $\{1, 2, 4\}$ & $\{5, 9\}$/- \\
            \multicolumn{2}{c}{}                                                                        & Layer 2 & $S_1 S_2 / 2$ & $\{8, 32, 64\}$                                                    & $\{2, 4, 8\}$ & $\{5, 9\}$/- \\ \midrule
            \multicolumn{2}{c}{\begin{tabular}[c]{@{}c@{}}KAN-1/\\ FIR\end{tabular}}                    & Layer 1 & 1                   & \begin{tabular}[c]{@{}c@{}}$\{21, 51,$\\ $121, 321\}$\end{tabular} & 2             & 17/-         \\ \bottomrule
        \end{tabular}
        \label{tab:search_space}
    \end{minipage}
    \hfill
    \begin{minipage}[c]{0.63\textwidth}
        \centering
        \input{figures/paretos}
    \end{minipage}
    
    \begin{minipage}[t]{0.36\textwidth}
    \end{minipage}
    \hfill
    \begin{minipage}[t]{0.63\textwidth}
        \centering
        \captionsetup{width=\linewidth}
        \captionof{figure}{Pareto fronts for unpruned (a) and pruned (b) networks at $\text{ROP}=-2$\,dBm and 2.2\,km (C-band) considering the maximum over all training sequences of the mean $\overline{\text{BER}}$ taken at the last ten training iterations.}
        \label{fig:paretos}
    \end{minipage}
\end{figure}

\vspace{-1em}
\section{Conclusion}

We showed that the novel \acp{KAN} are able to outperform \acp{CNN} as equalizers in \acp{PON} at the same complexity, when compared in terms of \ac{rvms}, while successfully compensating for the non-linearities in high \acp{ROP} also at low computational complexity.

\vspace*{2ex}
\footnotesize 
\noindent{\bf{Acknowledgements}}:
This work has received funding from the European Research Council (ERC)
under the European Union’s Horizon 2020 research and innovation program
(grant agreement No. 101001899).

\end{document}